\documentclass{article}

\usepackage{titling}
\usepackage[centertags]{amsmath}
\usepackage{pbox,chicago,fancyhdr}
\usepackage{amssymb,amsfonts,eucal,oldgerm,textcomp}
\usepackage[text={6in,8.25in},centering]{geometry}

\usepackage[colorlinks=true,citecolor=red,linkcolor=blue,urlcolor=blue,pdftex]{hyperref}

\newcommand{\skipline}[1][1]{\vspace*{#1\baselineskip}}

\newcommand{\smidge}{\hspace*{.1em}}

\newcommand{\half}{\frac{1}{2}}

\title{Kinematics, Dynamics, and the Structure of Physical
  Theory\thanks{This paper has been submitted to \emph{Philosophy of
      Science}, Mar.\@ 2016.  I thank Chris Pincock for detailed
    comments, insightful suggestions and hard questions on an earlier
    draft of a manuscript of which this paper is a fragment, and for
    many enjoyable, illuminating conversations about these things in
    general.  I thank Adam Caulton, Bill Demopoulos, and Sebastian
    Lutz for enjoyable and helpful conversations on the structure and
    semantics of theories in general.  Finally, I thank Howard Stein
    for many fruitful and delightful conversations over many years
    touching on all sorts of matters related to the issues I address
    here in particular, and, in general, for more than I can well say.
    This paper owes a clear and debt to several of his papers,
    especially Stein~(1992, 1994, 2004).}}

\author{Erik Curiel\thanks{\textbf{Author's address}: Munich Center
    for Mathematical Philosophy, Ludwigstra{\ss}e 31,
    Ludwig-Maximilians-Universit\"at, 80539 M\"unchen, Deutschland;
    \textbf{email}: \href{mailto:erik@strangebeautiful.com}
    {\texttt{erik@strangebeautiful.com}}}}

\predate{\begin{center} \large}
\postdate{\end{center} \skipline[-2.75]}

\date{}

\begin{document}

\thispagestyle{empty}
\maketitle

\nocite{stein-struct-know,stein-carnap-not-wrong,stein-enterprise}

\begin{quote}
  \begin{center}
    \textbf{ABSTRACT}
  \end{center}  

  Every physical theory has (at least) two different forms of
  mathematical equations to represent its target systems: the
  dynamical (equations of motion) and the kinematical (kinematical
  constraints).  Kinematical constraints are differentiated from
  equations of motion by the fact that their particular form is fixed
  once and for all, irrespective of the interactions the system enters
  into.  By contrast, the particular form of a system's equations of
  motion depends essentially on the particular interaction the system
  enters into.  All contemporary accounts of the structure and
  semantics of physical theory treat dynamics, \emph{i}.\emph{e}., the
  equations of motion, as the most important feature of a theory for
  the purposes of its philosophical analysis.  I argue to the contrary
  that it is the kinematical constraints that determine the structure
  and empirical content of a physical theory in the most important
  ways: they function as necessary preconditions for the appropriate
  application of the theory; they differentiate types of physical
  systems; they are necessary for the equations of motion to be well
  posed or even just cogent; and they guide the experimentalist in the
  design of tools for measurement and observation.  It is thus
  satisfaction of the kinematical constraints that renders meaning to
  those terms representing a system's physical quantities in the first
  place, even before one can ask whether or not the system satisfies
  the theory's equations of motion.
\end{quote}

\newpage

\tableofcontents

\section{Introduction}
\label{sec:intro}

Every physical theory has (at least) two different forms of
mathematical equations to represent its target systems: the dynamical
(equations of motion) and the kinematical (kinematical constraints).
Since at least the seminal work of
Suppes~\citeyear{suppes-mngs-uses-mods,suppes-mods-data}, contemporary
investigation and analysis of the structure and semantics of physical
theories has focused on the character and role of a theory's equations
of motion.  In particular, the family of solutions to the equations of
motion, and the models those solutions allow one to construct, have
taken pride of place in determining the structure and semantics of a
theory.  This is true whether one hews to the semantic view of
theories \cite{suppe-srch-underst-sci-thry,fraassen-sci-image} or the
Best-Systems picture \cite{cohen-callender-better-best} or a semantics
based on possible worlds \cite{lewis-def-theor-terms}, or one is a
neo-Carnapian \cite{demopoulos-log-phil-leg}, or a structuralist
\cite{costa-french-models-sci-reason}, or a neo-Kantian
\cite{friedman-dyns-reason}, or an inferentialist
\citeN{suarez-infer-conc-rep}, or any other of the contemporary
popular accounts of scientific theory.  Only a fool or a philosopher
would deny that the dynamics of a theory plays a central role of
fundamental importance in a proper accounting of its structure and
semantics.  I believe, however, that focus on the dynamics to the
exclusion of other fundamental structures theories possess can give at
best only a partial picture of a theory, and in many if not most cases
a distorted, misleading and even wildly inaccurate one.

I argue that it is exactly satisfaction of the kinematical
constraints---fixed, unchanging relations of constraint among the
possible values of a system's physical quantities---that ground the
idea of the individual state of a system as represented by a given
theory.  If the individual quantities a theory attributes to a system
do not stand in the minimal relations to each other required by the
theory, then the idea of a state as representing that kind of system
disintegrates, and without the idea of an individual state of a
system, one can do nothing in the theory to try to represent the
system.  \emph{A fortiori}, if the kinematical constraints are not
satisfied, one has no grounds for believing that the system at hand is
one of the type the theory treats.  It is thus those constraints that
differentiate types of physical systems, and not their dynamics.
Kinematical constraints, therefore, also function as necessary
preconditions for the appropriate application of the theory in the
first place, before one can even ask whether a given system the theory
purportedly treats satisfies its equations of motion.  Indeed, they
are necessary for the equations of motion to be well posed or even
just cogent.  Finally, they, and not the equations of motion, guide
the experimentalist in the design of tools for measurement and
observation.  It is thus satisfaction of the kinematical constraints
that renders meaning to those terms representing a system's physical
quantities in the first place, even before one can ask whether or not
the system satisfies the theory's equations of motion.

\section{Kinematics and Dynamics}
\label{sec:kins-dyns}

It is often useful when contemplating a physical theory to distinguish
its kinematical from its dynamical components.  I begin with a general
account of this.

The difference between the kinematic and the dynamic manifests itself
first in the family of quantities a theory ascribes to a type of
system.  On the one hand, there are the quantities that can vary with
time and place while the system remains otherwise individually the
same; these are the dynamic quantities (position, velocity, angular
momentum, shear-stress, electric current, \ldots).  On the other,
there are the quantities that one assumes, for the sake of argument
and investigation, remain constant as the system dynamically evolves,
on pain of the system's alteration \emph{in specie}; these are the
kinematic quantities (Hooke's constant, electrical resistance, shear
viscosity, thermoconductivity, index of refraction, \ldots).  This
classification belongs to kinematics.  

A state of a system is the aggregation of the values of its physically
significant properties at an instant; it is represented by a
proposition encapsulating all that can be known of the system
physically, at least so far as the theoretical and experimental
resources one relies on are concerned.  If one can distinguish the
values of the properties of the system at one time from those at
another time by the available resources, then the system is in a state
at the first time different from that at the second.  A state,
therefore, can be thought of as a set of the values of quantities that
jointly suffice for the identification of the species of the system
and for its individuation at a moment.  As such, the state is the most
fundamental unit of theoretical representation of a system \emph{as} a
unified system, rather than just as (say) a bunch of random, unrelated
properties associated with a spatiotemporal region.  The
characterization of a system's state belongs to kinematics.  Every
known physical system has the property that at least some of its
quantities almost always change in value as time passes, which is to
say, the system in general occupies different states at different
moments of time.  The collection of states it serially occupies during
an interval of time forms a \emph{dynamical evolution} (or just
`possible evolution').  The characterization of possible evolutions
belongs to dynamics.

Roughly speaking, then, kinematics comprises what one needs to know in
order to fix the type of system at issue (is it a viscous fluid?  an
electromagnetic field?), and to give a complete description of its
state at a single moment---complete, that is, with respect to the
theory at issue, \emph{i}.\emph{e}., a consistent ascription of values
to all the quantities it bears that are treated by a model of it in
the theory.  Dynamics comprises what one needs to know in order to
individuate a system and to describe its behavior over time, in order
to conclude, for example, that one's model represents this system
right here by the determination of the values that a particular set of
its quantities respectively takes over the next 5 minutes, given both
its state at the initial moment and the state of its environment (the
forces, if any, it is subject to, or the interactions it enter into)
at that moment and over the course of those 5 minutes.

Kinematics does more than classify the quantities of a type of
physical system into the kinematic and the dynamic.  It also imposes
fixed, unchanging relations of constraint among their possible values,
both constraints that must hold at a single instant and those that
must hold over the course of any of the system's possible evolutions.
More precisely, there are two kinds of kinematical constraints a
theory may comprise, the \emph{local} and the \emph{global}.  A local
constraint involves only quantities that can be attributed to a single
state of the system, such as position; a global one involves a
quantity that cannot be attributed to any single state of the system,
such as the period of an orbiting body.\footnote{There is a subtlety
  here.  Any kinematical constraints that involve derivatives depend,
  strictly speaking, on values of quantities at more than one state,
  even for local constraints; some global constraints, moreover, can
  be formulated by laying down conditions that must hold at individual
  states (\emph{e}.\emph{g}., that a Newtonian orbit be an ellipse can
  be formulated as a constraint on the value of the spatial derivative
  at every point of the orbit, or on the sum of the distances from the
  foci at each point); this seems superficially similar to some local
  ones, \emph{e}.\emph{g}., conservation of angular momentum, which
  can also be formulated as a relation among derivatives at a point.
  Whether a constraint, then, is global or local, depends in part on
  whether one can formulate the condition over arbitrarily short
  periods of a possible evolution, which one can for conservation of
  angular momentum (the system satisfies angular momentum, say, during
  one part of an evolution but not another), but not for whether a
  planetary orbit is an ellipse (where, by definition, one must wait
  an entire orbital period before one can say the condition is
  satisfied or not).}  Examples of kinematical constraints:
\begin{itemize}
  \setlength{\itemsep}{0em}
    \item Hooke's constant $k$ has physical dimension
  $\displaystyle \frac{m}{t^2}$ (local)
    \item the shear-stress tensor is symmetric in Navier-Stokes
  theory, $\sigma_{ab} = \sigma_{(ab)}$ (local)
    \item Kepler's Harmonic Law,
  $\displaystyle \frac{a^3} {T^2} = M$ (global)
    \item stress-energy tensor is covariantly divergence-free in
  general relativity, $\nabla^n \smidge T_{na} = 0$ (local)
    \item the Heisenberg uncertainty principle,
  $\Delta x \smidge \Delta p \ge \half \hbar$ (local)
\end{itemize}

I shall spend most of the rest of the paper discussing kinematics (and
dynamics mostly by way of contrast).  Although there is much more to
say about the dynamical structure of a physical theory, for the
purposes of this paper I must rest content with remarking that it
includes in general a rich and deep lode of topological, geometrical,
analytical and algebraic structures on the space of states that in
particular encode relations among entire classes of dynamic
evolutions; those relations often take in part the form of a set of
partial-differential equations expressed in terms of the kinematic and
dynamic quantities, the equations of motion, the solutions to which
represent the totality of the system's dynamical evolutions starting
from all kinematically possible initial states.  The canonical example
is Newton's Second Law: a Newtonian body accelerates in direct, fixed
proportion to the net total force applied to it, the ratio of the
acceleration to the total force being the kinematic quantity known as
the body's inertial mass.

\section{Kinematical Constraints}
\label{sec:kinematical-constraints}

Kinematical constraints are differentiated from equations of motion by
the fact that the particular, concrete form of a kinematical
constraint is fixed once and for all, irrespective of the interactions
the system may enter into with other systems (such as a measuring
apparatus in the laboratory).  By contrast, the particular, concrete
form of a system's equations of motion depends essentially on the
particular interaction (if any) the system enters into with another
system in its environment---\emph{e}.\emph{g}., what external forces,
if any, act on the system.  

The difference between a kinematical constraint and an equation of
motion comes out clearly in Newton's Second Law, written out
explicitly as two coupled first-order differential equations.
\begin{equation*}
  \dot{\mathbf{x}} = \mathbf{v}
\end{equation*}
\begin{center}
  \skipline[-.5] (always the same: kinematical constraint)
\end{center}
\begin{center}
  versus
\end{center}
\begin{equation*}
  \dot{\mathbf{v}} = \mathbf{F}/m
\end{equation*}
\begin{center}
  \skipline[-.5] (the concrete form of $\mathbf{F}$ depends on
  environment, forces: equation of motion)
\end{center}
For a more interesting example, consider the Maxwell equations.
According to this characterization, the first two,
\begin{equation}
  \label{eqn:maxwell-constraints}
  \begin{split}
    \boldsymbol{\nabla} \boldsymbol{\cdot} \mathbf{B} &= 0 \\
    \dot{\mathbf{B}} &= -\boldsymbol{\nabla} \times \mathbf{E} \\
  \end{split}
\end{equation}
those governing the magnetic components $\mathbf{B}$ of the
electromagnetic field, are both local kinematical constraints.  They
are kinematical constraints and not equations of motion because
neither changes form no matter the environment the electromagnetic
field evolves in (ignoring the possibility of magnetic monopoles).
Indeed, even though one of the equations includes the time-derivative
of another quantity, making it look like an equation of motion, I
claim that from a physical point of view one must think of them both
as kinematical constraints.  The crux of the matter is that the
electromagnetic field couples with other systems only by way of their
manifestation of electric charge $\rho$ or current $\mathbf{j}$, but
those quantities when present change the form only of the other two
Maxwell equations,
\begin{equation}
  \label{eqn:maxwell-eoms}
  \begin{split}
    \boldsymbol{\nabla} \boldsymbol{\cdot} \mathbf{E} &= \rho \\
    \dot{\mathbf{E}} &= \mathbf{j} - \boldsymbol{\nabla} \times
    \mathbf{B} \\
  \end{split}
\end{equation}
those governing the electric components $\mathbf{E}$ of the
electromagnetic field.  In effect, the difference between the two
pairs of relations shows that, in a precise sense, the magnetic field
couples directly with no physical quantity of any other system in that
the presence of electric charges and currents does not alter the form
of its two defining equations.  (The magnetic field does couple to
electric current ``to second order'' by way of the second of
equations~\eqref{eqn:maxwell-eoms}, whence Amp\`ere's Law.)  Thus the
form of equations~\eqref{eqn:maxwell-constraints} does not depend on
the particular dynamical evolution the system manifests at any given
time.  Nonetheless, not just any old thing counts as a magnetic field
no matter how it evolves and no matter what relations hold among its
quantities at different points; only those things that behave like
magnetic fields can be magnetic fields, which in this case means the
identical satisfaction of the first two Maxwell equations.

\section{Roles in Theory}
\label{sec:roles-theory}

Theories do not predict kinematical constraints; they demand them.  I
take a prediction to be something that a theory, while appropriately
modeling a system, can still get wrong.  Newtonian mechanics, then,
does not predict that the kinematical velocity of a Newtonian body
equal the temporal rate of change of its position; rather it requires
it as a precondition for its own applicability.  It can't ``get it
wrong''.  If the kinematical constraints demanded by a theory do not
hold for a family of phenomena, that theory cannot treat it, for the
system is of a type beyond the theory's scope.  By contrast, if the
equations of motion are not satisfied, that may tell one only that one
has not taken all ambient forces on the system (couplings with its
environment) into account; it need not imply that one is dealing with
an entirely different form of system.  Even in principle, one can
never entirely rule out the mere possibility that the equations of
motion are inaccurate only because there is a force one does not know
how to account for, not because the system is not accurately treated
by those equations of motion.  This can never happen with a
kinematical constraint.  It is either satisfied, to the appropriate
and required level of accuracy given the measuring techniques
available and the state of the system and its environment, or it is
not.  This is a serious difference in physical significance among the
types of proposition a theory contains, which, among other things,
should be reflected in the way an account of semantics assigns
significance to the theory's structural elements.

Indeed, satisfaction of kinematical constraints is required for the
equations of motion of a theory to be well posed or even just cogent.
The initial-value formulation of the Navier-Stokes equations, for
example, is well set (in the sense of Hadamard) only if the
shear-stress tensor is symmetric and the heat flux is orthogonal to
fluid flow, both kinematical constraints
\cite{lamb-hydro,landau-lifschitz-fluid}.  One cannot even formulate
Newton's Second Law if velocity is not the first temporal derivative
of position.  More generally, in a sense one can make precise
([*** citation redacted for anonymity ***]), if the kinematical
constraints of Lagrangian mechanics are not satisfied
($\mathbf{v} = \mathbf{\dot{q}}$), then one cannot formulate the
Euler-Lagrange equation; and similarly, if the kinematical constraints
of Hamiltonian mechanics are not satisfied (the $\mathbf{p}$s and
$\mathbf{q}$s do not satisfy the canonical Poisson-bracket
relations\footnote{$(q_i, \, p_j)$ satisfy the canonical
  Poisson-backet relations if
\begin{equation}
  \label{eq:can-pois-brkt}
  \begin{split}
    \{q_i, \, q_j\} &= 0 \\
    \{q_i, \, p_j\} &= \delta_{ij} \\
    \{p_i, \, p_j\} &= 0
  \end{split}
\end{equation}
where $\delta_{ij}$ is the Kronecker delta symbol, which equals 1 for
$i = j$ and 0 otherwise.}), then one cannot formulate Hamilton's
equation.  Thus satisfaction of the
kinematical constraints is required as a precondition for the
appropriate application of a theory in modeling a kind of system, and
so the kinematical constraints in fact function in that precise sense
as \emph{a priori} constitutive components of a physical theory.

This is not true of the dynamical relations the theory posits.  A
theory may appropriately treat a family of phenomena even when it does
not model the dynamical behavior of all members of the family to any
prescribed degree of accuracy, \emph{i}.\emph{e}., even when the
equations of motion are not satisfied in any reasonable sense (and
thus when, according to the standard conception of semantics, the
schematic representations of those phenomena cannot contribute to the
semantic content of the terms occurring in those representations).  A
theory, however, can and does tell us much about the character and
nature of physical systems for which it does not give accurate
representations, systems, in other words, it cannot soundly represent
in totality, cannot be true of, and so systems that, according to all
the standard contemporary accounts of theory structure and semantics,
the theory should have nothing to say about at all.  If a system's
behavior is not accurately captured by a theory's equations of motion,
then that system cannot, \emph{e}.\emph{g}., be represented by a
Tarskian model constructed from a solution to the equations of motion;
it is thus, according to the semantic view of theories, for instance,
not even a candidate for contributing to the semantic content of the
theory's theoretical terms, \emph{inter alia}.  In fact, though, such
systems can still be appropriately represented by that theory in a
precise and important sense, even though the equations of motion are
not satisfied, so long as the kinematical constraints are.

Consider the example of a representation of a body of liquid as
provided by the classical theory of fluid mechanics, Navier-Stokes
theory.  When the liquid is not too viscous, is in a state near
hydrodynamical and thermodynamical equilibrium, and the level of
precision and accuracy one demands of the representation is not at too
fine a spatiotemporal scale, then the classical theory yields
excellent models of the liquid's behavior over a wide range of states
and environments.  When the state of the liquid, say, begins to
approach turbulence, the representation the theory provides begins to
break down.  It does so, however, in a subtle way, one that cannot be
wholly accounted for by adverting merely to the fact that the theory
becomes predictively inaccurate.  In particular, there is a regime in
which the theory's dynamical equations of motion no longer provide
accurate predictions by any reasonable measure, and yet all the
quantities the theory attributes to the liquid (\emph{e}.\emph{g}.,
shear viscosity, mass density, hydrostatic pressure, shear-stress,
\emph{et al}.\@) will still be well defined, and all the kinematical
constraints the theory jointly imposes on those quantities
(\emph{e}.\emph{g}., the constancy of shear viscosity, the continuity
of mass-density, the conservation of energy, the symmetry of the shear
tensor, \emph{etc}.), will still be satisfied
\cite{monin-yaglom-stat-fld-mech-turb}.  Call it \emph{the regime of
  kinematical propriety}.  In a strong sense, then, the theory can
still provide a meaningful---and appropriate---model of the liquid
even though that model is not adequately accurate in all its
predictions.  This sort of situation, where the theory's dynamics are
no longer adequate but its kinematics are still appropriate, shapes
and provides at least part of the physical meaning of terms like `mass
density' and `shear'---physical meaning that \emph{ipso facto} cannot
be captured by a semantics that grounds meaning on the dynamics of the
theory, and in particular by one that relies wholly or even in large
on part on the family of solutions to the theory's equations of
motion.

More precisely, then, a view about the structure and semantics of
physical theory based ultimately on dynamics is inadequate for (at
least) two reasons.  First, it does not allow us, within the scope of
the theory itself, to understand why such models are not sound even
though all the quantities the theory attributes to the system are well
defined and the values of those quantities jointly satisfy all
kinematical constraints the theory requires.  Second, we miss
something fundamental about the meaning of various theoretical terms
by rejecting such models out of hand merely on the grounds of their
inaccuracy.  It is surely part of the semantics of the term
`hydrostatic pressure', \emph{e}.\emph{g}., that its definition as a
physical quantity treated by classical fluid mechanics breaks down
when the fluid approaches turbulence; because, however, the theory's
equations of motion stop being accurate long before, in a precise
sense, the quantity loses definition in the theory and long before the
kinematical constraints of the theory stop being satisfied, any
account of the structure of theories and their semantics that rejects
the inaccurate models in which the term still is well defined will not
be able to account for that part of the term's meaning.  Thus, an
adequate account of physical theory must be grounded on notions
derived from relations in some sense prior to the theory's
representations of the dynamical behavior of the physical systems it
treats, relations that govern the propriety of the theory's
representational resources for modeling the system at issue.  These
are the the theory's kinematical constraints.

One may think that this discussion about how, where and when theories
breakdown more properly belongs to pragmatics (in the sense of
semiotic theory) than to semantics.  That is not so.  A system of
formal semantics that would ground itself in the family of possible
physical systems for which it provides sound models cannot even get
started until that family is demarcated.  But that is exactly to
require an investigation of the boundary of the theory's regime of
kinematical propriety, which is thus logically and conceptually prior
to any such system of semantics.

In order to be able to formulate and evaluate any kinematical
constraint, of course, the quantities themselves in the terms of which
the constraints are formulated must be well defined in the theory.
For this to be the case, it is necessary that one be able to formulate
the local kinematical constraints and verify that they hold.  Without
the satisfaction of the local kinematical constraints, the entire idea
of the individual state of a system as represented by that theory
disintegrates---individual quantities do not stand in the minimal
relations to each other required by the theory---and without the idea
of a state of a system, one can do nothing in the theory to try to
treat the system.

More to the point, if the local kinematical constraints are not
satisfied, one has no grounds for believing that the system at hand is
one of the type the theory treats.  Many different kinds of system,
for example, have shear and stress---Navier-Stokes fluids, elastic
solids, ionically charged plasmas, electromagnetic fields, \emph{et
  al}.  To say that a system has a quantity represented by a
shear-stress tensor is not to have said very much.  One must also
know, among other things, whether the shear-stress tensor must be
symmetric, or divergence-free, or stand in a fixed algebraic relation
to another of the system's quantities such as heat flux, and so on.
Each such possible condition is a kinematical constraint; and each
different type of system that has a quantity appropriately represented
by a shear-stress tensor will impose different constraints on that
tensor.  It is those constraints that differentiate types of physical
systems, and not their dynamics.  Think of all the kinds of systems
whose dynamics obey the equation of a simple harmonic oscillator
(pendulum, spring, vibrating string, electrical circuit, orbiting
planet, trapped quantum particle, \ldots)---without question what
differentiates them cannot be the form of their dynamics.  It is only
the form and content of the kinematical constraints one demands be
obeyed by the quantities entering into the equations of motion.  In
this sense, then, the kinematical constraints are constitutive of the
type of system the theory treats.

The same considerations show that kinematical constraints are, in a
precise sense, analytic: they are made true solely by the meanings of
the terms \emph{in the context of the theory}.  In that sense they are
like L-sentences in a Carnapian framework \cite{carnap-mean-nec}.
Unlike L-sentences, however, they have non-trivial semantic content,
for the constraints they impose on physical system are non-trivial.
Not all types of physical system will satisfy them, \emph{viz}., those
systems not appropriately represented by the theory.

Finally, it is the kinematical constraints, not the equations of
motion, that guide the experimentalist in the design of instruments
for probing and measuring the quantities the theory attributes to the
systems it treats.  An instrument that is to measure velocity, for
instance, must be sensitive to differences in spatial location at ever
smaller measured temporal intervals.  It does not care about how the
system accelerates, \emph{i}.\emph{e}., about its dynamics.
Similarly, an instrument that would measure shear-stress of a
Navier-Stokes fluid must conform to the equality of pressure and
reversed sense of shear across imaginary surfaces in fluid that is
represented by the symmetry of the shear-stress tensor.  Again, the
instrument need not care at all about the dynamics of the fluid to
measure the shear-stress.  In this way, they provide the foundation
for the operationalization of the meaning of theoretical
terms.\footnote{If one likes, one can take this as a way to make
  precise the sense in which experiments are ``theory laden'', and why
  that is irrelevant for the capacity of experiments to provide
  independent confirmation and refutation of theories: the equations
  of motion in general play no role in the design of experimental
  instruments, but it is, in general, only the equations of motion we
  test in experiments.}

To summarize, then, the roles that kinematical constraints play in
physical theory:
\begin{enumerate}
    \item they govern the propriety of theory in representing systems
  in the first place, \emph{i}.\emph{e}., they serve as preconditions
  of applicability
    \item they characterize the physical nature of systems the theory
  treats, \emph{i}.\emph{e}., that constitutive of the kind of system
  the theory treats
    \item they guarantee the cogency and good behavior of the
  dynamics, in so far as that can be guaranteed (by ensuring the well
  posedness of the initial-value formulation of the equations of
  motion, or by ensuring that the equations of motion are cogent as
  equations in the first place)
    \item they provide guidance in the design of tools for measurement
  and observation, and so provide the empirical ground for the meaning
  of theoretical terms
\end{enumerate}
The equations of motion play none of these roles.

Before concluding the paper in the next section, it will be
instructive to compare the way I have characterized kinematical
constraints with the manifestly (and superficially) similar ideas in
Neo-Kantian accounts of the structure and semantics of physical
theory, such as those of \citeN{reichenbach-rel-apriori-know} and
\citeN{friedman-dyns-reason}.  They postulate a relativized \emph{a
  priori}, which also is in some sense constitutive of the kinds of
systems treated by a theory, and which function in some sense as
preconditions for the applicability of a theory.  Kinematical
constraints, on my conception, do have some similarities to that idea,
but they have deep differences as well.
\begin{enumerate}
    \item First and foremost, kinematical constraints are part of the
  theory itself, not supra-theoretical principles.
    \item \emph{Contra} several of the Reichenbachian examples of
  relativized \emph{a priori} principles, such as that of genidentity
  \cite{padovani-rel-rel-apriori}, kinematical constraints have true
  physical content, not just formal character, in the sense that
  direct measurement can verify whether they hold or not of a given
  system.
    \item One needs the satisfaction of kinematical constraints,
  \emph{as experimentally verified}, in order to apply the theory
  appropriately in the most full-blooded sense, that of characterizing
  systems and making predictions about them.
\end{enumerate}
With regard to the last point, \citeN[p.~71]{friedman-dyns-reason}
does say, ``The role of constitutively \emph{a priori} principles is
to provide the necessary framework in which the testing of properly
empirical laws is possible.''  Nonetheless, \emph{a priori} principles
on his conception are not amenable to direct experimental verification
in the same way as kinematical constraints.  Kinematical constraints,
as opposed to the kind of \emph{a priori} principles he characterizes,
appear already as part of the theory itself, rigorously and precisely
formulated---and so amenable to direct experimental testing---not as
imprecise, loose and supra-theoretic adjuncts to the theory.

\section{Bearing on Semantics}
\label{sec:bearing-semantics}

To accept a theory is, at a minimum, to accept its analytic or \emph{a
  priori} propositions as true---as necessarily true in the context of
the framework.  To accept Newtonian mechanics is to accept that
$\mathbf{v} = \mathbf{\dot{x}}$ and that
$m \mathbf{\dot{v}} = \mathbf{F}$.  It is not to accept that the
gravitational force is $\displaystyle G \frac{m_1 m_2} {r^2}$, nor to
accept that the net force on this body right here, right now, is 5
Newtons.  One requires a semantics of frameworks that allows one to
demarcate that class of propositions, the ones necessarily true in the
context of the framework.  One cannot know them as analytic if given
the framework only as a formal structure, or if one uses a semantics
such as a Tarskian one that treats all propositions as semantically on
par with each other.  The apriority of the propositions must come as
part of the semantic interpretation of the framework itself.

One may say that a theory has \emph{propriety of representation} for a
system when the system satisfies its kinematical constraints, for
their satisfaction is semantically prior to the satisfaction of the
equations of motion (\S\ref{sec:roles-theory}).  It therefore seems
promising to attempt to base a semantics for physical theory on this
idea:
\begin{quote}
  We know the meaning of a theory when we know the conditions under
  which the kinematical constraints hold, \emph{i}.\emph{e}., when the
  the theory has propriety in representation.
\end{quote}
To know the meaning of a theory, therefore, cannot be to know the set
of ``possible worlds'' the solutions to the theory's equations of
motion represents.  It is rather to know the conditions under which it
is sensical to investigate the formulation of possible conditions of
the theory's truth, \emph{i}.\emph{e}., the satisfaction of its
equations of motion, for this can be done only in so far as one
already knows what systems the theory represents with propriety.


\begin{thebibliography}{}

\bibitem[\protect\citeauthoryear{Carnap}{Carnap}{1956}]{carnap-mean-nec}
Carnap, R. (1956).
\newblock {\em Meaning and Necessity: A Study in Semantics and Modal Logic\/}
  (Second ed.).
\newblock Chicago: The University of Chicago Press.
\newblock First edition published 1947. Midway Reprint Edition printed 1988.

\bibitem[\protect\citeauthoryear{Cohen and Callender}{Cohen and
  Callender}{2009}]{cohen-callender-better-best}
Cohen, J. and C.~Callender (2009).
\newblock A better best system account of lawhood.
\newblock {\em Philosophical Studies\/}~{\em 145\/}(1), 1--34.
\newblock \href{http://dx.doi.org/10.1007/s11098-009-9389-3}
  {doi:10.1007/s11098-009-9389-3}.


\bibitem[\protect\citeauthoryear{da~Costa and French}{da~Costa and
  French}{2005}]{costa-french-models-sci-reason}
da~Costa, N. and S.~French (2005).
\newblock {\em Science and Partial Truth: A Unitary Approach to Models and
  Scientific Reasoning}.
\newblock Oxford: Oxford University Press.

\bibitem[\protect\citeauthoryear{Demopoulos}{Demopoulos}{2013}]{demopoulos-log-phil-leg}
Demopoulos, W. (2013).
\newblock {\em Logicism and Its Philosophical Legacy}.
\newblock Cambridge: Cambridge University Press.

\bibitem[\protect\citeauthoryear{Fraassen}{Fraassen}{1980}]{fraassen-sci-image}
Fraassen, B.~v. (1980).
\newblock {\em The Scientific Image}.
\newblock Oxford: Oxford University Press.

\bibitem[\protect\citeauthoryear{Friedman}{Friedman}{2001}]{friedman-dyns-reason}
Friedman, M. (2001).
\newblock {\em The Dynamics of Reason}.
\newblock Stanford, CA: CSLI Publications.
\newblock Delivered as the 1999 Kant Lectures at Stanford University.

\bibitem[\protect\citeauthoryear{Lamb}{Lamb}{1932}]{lamb-hydro}
Lamb, H. (1932).
\newblock {\em Hydrodynamics\/} (sixth ed.).
\newblock New York: Dover Publications.
\newblock The 1945 Dover reprint of the 1932 edition published by the Cambridge
  University Press.

\bibitem[\protect\citeauthoryear{Landau and Lifschitz}{Landau and
  Lifschitz}{1975}]{landau-lifschitz-fluid}
Landau, L. and E.~Lifschitz (1975).
\newblock {\em Fluid Mechanics\/} (Second ed.).
\newblock Oxford: Pergamon Press.
\newblock An expanded, revised edition of the original 1959 edition. Translated
  from the Russian by J. Sykes and W. Reid.

\bibitem[\protect\citeauthoryear{Lewis}{Lewis}{1970}]{lewis-def-theor-terms}
Lewis, D. (1970, July).
\newblock How to define theoretical terms.
\newblock {\em The Journal of Philosophy\/}~{\em 67\/}(13), 427--446.
\newblock Stable URL: \url{http://www.jstor.org/stable/2023861}.

\bibitem[\protect\citeauthoryear{Monin and Yaglom}{Monin and
  Yaglom}{1971}]{monin-yaglom-stat-fld-mech-turb}
Monin, A. and A.~Yaglom (1971).
\newblock {\em Statistical Fluid Mechanics: {T}he Mechanics of Turbulence},
  Volume~1.
\newblock Mineola, NY: Dover Publications, Inc.
\newblock A 2007 edition of the 1971 English translation published by MIT
  Press, translated by Scripta Technica, Inc. Originally published in 1965 by
  Nauka Press, Moscow, as \emph{Statisticheskaya gidromekhanika---Mekhanika
  Turbulentnosti}.

\bibitem[\protect\citeauthoryear{Padovani}{Padovani}{2011}]{padovani-rel-rel-apriori}
Padovani, F. (2011, July).
\newblock Relativizing the relativized a priori: {R}eichenbach's axioms of
  coordination divided.
\newblock {\em Synthese\/}~{\em 181\/}(1), 41--62.
\newblock \href{http://dx.doi.org/10.1007/s11229-009-9590-0}
  {doi:10.1007/s11229-009-9590-0}.

\bibitem[\protect\citeauthoryear{Reichenbach}{Reichenbach}{1965}]{reichenbach-rel-apriori-know}
Reichenbach, H. (1965).
\newblock {\em The Theory of Relativity and A Priori Knowledge}.
\newblock Berkeley, CA: University of California Press.

\bibitem[\protect\citeauthoryear{Stein}{Stein}{1992}]{stein-carnap-not-wrong}
Stein, H. (1992).
\newblock Was {Carnap} entirely wrong, after all?
\newblock {\em Synthese\/}~{\em 93}, 275--295.

\bibitem[\protect\citeauthoryear{Stein}{Stein}{1994}]{stein-struct-know}
Stein, H. (1994).
\newblock Some reflections on the structure of our knowledge in physics.
\newblock In D.~Prawitz, B.~Skyrms, and D.~Westerst{\aa}hl (Eds.), {\em Logic,
  Metholodogy and Philosophy of Science}, Proceedings of the Ninth
  International Congress of Logic, Methodology and Philosophy of Science, pp.\
  633--55. New York: Elsevier Science B.V.

\bibitem[\protect\citeauthoryear{Stein}{Stein}{2004}]{stein-enterprise}
Stein, H. (2004).
\newblock The enterprise of understanding and the enterprise of knowledge---for
  {Isaac Levi}'s seventieth birthday.
\newblock {\em Synthese\/}~{\em 140}, 135--176.
\newblock I do not have access to the published version of Stein's paper, but
  rather only to a typed manuscript. All references to page numbers, therefore,
  do not correspond to those of the published version. The typed manuscript I
  have is 65 pages long, and the published version about 41. Multiplying the
  page numbers I give by $\frac{41}{65}$ and adding the result to 135 (the
  number of the first page in the published version) should give approximately
  the page number in the published version.

\bibitem[\protect\citeauthoryear{Su\'arez}{Su\'arez}{2004}]{suarez-infer-conc-rep}
Su\'arez, M. (2004, December).
\newblock An inferential conception of scientific representation.
\newblock {\em Philosophy of Science\/}~{\em 71\/}(5), 767--779.
\newblock \href{http://dx.doi.org/10.1086/421415} {doi:10.1086/421415}.

\bibitem[\protect\citeauthoryear{Suppe}{Suppe}{1974}]{suppe-srch-underst-sci-thry}
Suppe, F. (1974).
\newblock The search for philosophic understanding of scientific theories.
\newblock In F.~Suppe (Ed.), {\em The Structure of Scientific Theories}, pp.\
  3--254. Chicago: University of Illinois Press.

\bibitem[\protect\citeauthoryear{Suppes}{Suppes}{1960}]{suppes-mngs-uses-mods}
Suppes, P. (1960, September).
\newblock A comparison of the meaning and uses of models in mathematics and the
  empirical sciences.
\newblock {\em Synthese\/}~{\em 12\/}(2--3), 287--301.
\newblock \href{http://dx.doi.org/10.1007/BF00485107} {doi:10.1007/BF00485107}.

\bibitem[\protect\citeauthoryear{Suppes}{Suppes}{1962}]{suppes-mods-data}
Suppes, P. (1962).
\newblock Models of data.
\newblock In E.~Nagel, P.~Suppes, and A.~Tarski (Eds.), {\em Logic, Methodology
  and Philosophy of Science}, pp.\  252--261. Palo Alto, CA: Stanford
  University Press.

\end{thebibliography}
\end{document}